\documentclass[showpacs,twocolumn,prb]{revtex4-1}
\usepackage{amsfonts}
\usepackage{graphicx}
\usepackage{amsmath}
\usepackage{amssymb}

\begin{document}

\title{Magneto-optical conductivity in a topological insulator}
\author{Zhou Li$^1$}
\email{lizhou@univmail.cis.mcmaster.ca}
\author{J. P. Carbotte$^{1,2}$}
\email{carbotte@univmail.cis.mcmaster.ca}

\affiliation{$^1$ Department of Physics, McMaster University,
Hamilton, Ontario,
Canada,L8S 4M1 \\
$^2$ Canadian Institute for Advanced Research, Toronto, Ontario,
Canada M5G 1Z8}

\begin{abstract}
Adding a small subdominant quadratic in momentum term to a dominant
linear Dirac dispersion curve affects conduction and valence band
differently and leads to an hourglass-like structure for energy as a
function of momentum. This applies to the protected surface states
in topological insulators. The energies of the conduction and
valence band Landau levels are also different and this leads to the
splitting of optical absorption lines produced by the magnetic
field, which acquire a two peak structure. It also changes the peaks
in the imaginary part of the Hall conductivity into two distinct
contributions of opposite signs. The real part of the circularly
polarized optical conductivity however retains its single peak
structure but the peaks in right and left handedness case are
shifted in energy with respect to each other in contrast to the pure
Dirac case. The magnitude of the semiclassical cyclotron frequency
is significantly modified by the presence of a mass term as is its
variation with value of the chemical potential $\mu$. Its optical
spectral weight is found to decrease with increasing $\mu$ rather
than increase as it does in the pure Dirac limit.
\end{abstract}

\date{\today}
\maketitle

\section{Introduction}
Topological insulators are insulating \cite{Hasan,Qi1,Hsieh1,Chen}
in their bulk but have protected metallic surface states which
support helical Dirac fermions at the $\Gamma $ point of the
honeycomb lattice surface Brillouin zone. The direction of the in
plane electron spin is locked to be perpendicular to its
momentum with opposite spin winding in conduction and valence band \cite%
{Hsieh2,Jozwiak,Xu} as was confirmed by spin polarized angular
resolved photoemission spectroscopy. In contrast to the case of
graphene where the Dirac cones are nearly symmetric between
conduction and valence band, topological insulators rather exhibit
an hourglass shape \cite{Hsieh2,Jozwiak,Xu,Qi2,Hancock,Li3} with
valence band fanning out and merging with the bulk valence band. To
get the Fermi level to lie in the surface states between bottom of
the bulk conduction and top of bulk valence band requires care but
this can be done by doping. For example one can dope with $Sn$ in
$(Bi_{1-\delta }Sn_{\delta })_{2}Te_{3}$ (see
Ref.[\onlinecite{Chen}]) or $Ca$ in $Bi_{2-\delta }Ca_{\delta
}Se_{3}$.\cite{Hsieh2} The dynamics of charge carriers can be probed
by optics. The real part of the complex frequency dependent
longitudinal optical conductivity $\sigma (\omega )$ gives the
absorption as a function of photon energy. When applied to graphene
good agreement is found between theory and
experiment.\cite{Carbotte1,Carbotte2,Li,Onlita,Crasee,Nicol,Stauber}
The technique has also been applied to get information on the
surface states of topological insulators \cite{Hancock} and other
single layered materials such as $MoS_{2}$ \cite{Li12,Li13} and
silicene \cite{Stille}. Additional valuable information results when
a magnetic field is applied. This creates Landau levels (LL) and
incident photons can induce transition between these levels.\cite
{Carbotte3,Pound,Sadowski,Jiang,Tabert} Recently it has been applied
to the topological insulator $Bi_{0.91}Sb_{0.09}$. \cite{Schafga} In
this paper we consider in detail the magneto-optical conductivity of
a topological insulator with particular emphasis on the hourglass
shape of the Dirac cones. This comes from a quadratic in momentum
\cite{Ando} piece in the electron dispersion and is additional to
the usual Dirac part which gives a contribution to the energy linear
in momentum. It is well known that a quadratic alone gives LL
spacings proportional to the magnetic field $B$ while the linear
piece alone gives spacings which are drastically different going
instead as the square root of $B$. This has important implications
for the optical absorption when both parts are present in the
fermion dispersion curves as we will find here.

In section II we specify our Hamiltonian and provide solutions for
the energy eigenvalues and eigenfunctions under magnetic field. In
section III we introduce the formal expressions needed to compute
the magneto conductivity and provide numerical results. Section IV
deals with the semiclassical limit when the chemical potential is
much larger than the Landau level spacing. Conclusions and a summary
are found in section V.

\section{Some formalism}
\begin{figure}[tp]
\begin{center}
\includegraphics[height=2.0in,width=2.0in]{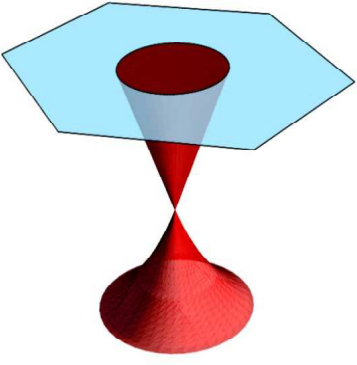}
\end{center}
\caption{(Color online) Schematic of Dirac cone centered about the
$\Gamma $ point of the honeycomb Brillouin zone of the surface
states in a topological insulator. The top cone is the conduction
band while the bottom cone which gives the figure a goblet or
hourglass shape is for the valence band. } \label{fig1}
\end{figure}

\begin{figure}[tp]
\begin{center}
\includegraphics[height=3.0in,width=3.0in]{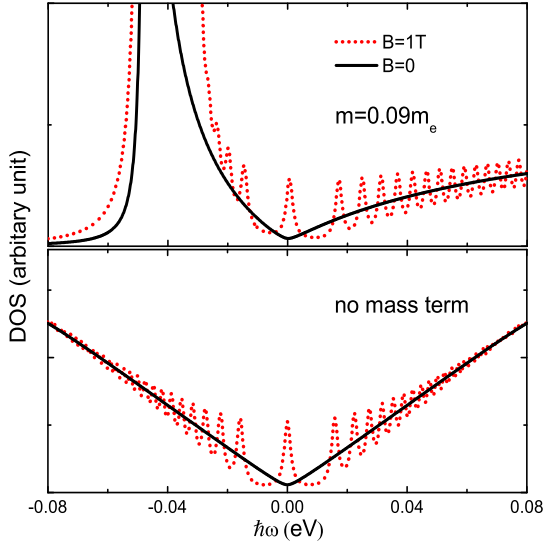}
\end{center}
\caption{(Color online) The density of state as a function of energy
$\hbar\omega$ in eV for a topological insulator with mass
$m=0.09m_{e}$ (top frame) and no mass term ($m=\infty $) (bottom
frame) shown for comparison. The magnetic field B=1 Tesla. The
dotted red curves include B while the solid black are the case of no
magnetic field (B=0) included for comparison. The parameters are for
$Bi_{2}Te_{3}$, $m=0.09m_e$ and $\alpha/\hbar=4.3\times10^{5}$m/s. }
\label{fig2}
\end{figure}

\begin{figure}[tp]
\begin{center}
\includegraphics[height=4in,width=3.2in]{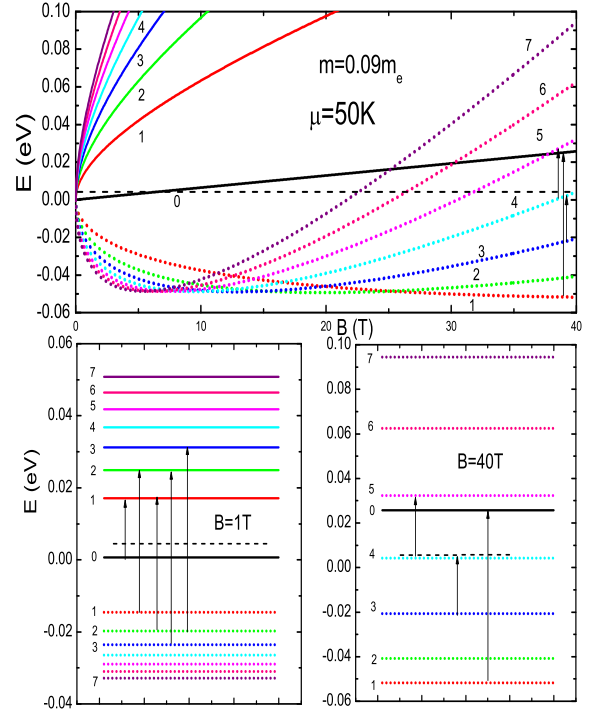}
\end{center}
\caption{(Color online) The top panel gives the energies $E_{N,s}$
of the Landau levels for various values of N (as labeled) as a
function of the magnetic field B. The conduction band levels are
solid curves and the valence band dotted curves. Here the quadratic
mass term has $m=0.09m_{e}$. The horizontal dashed line is the
chemical potential level $\mu=50K$. In the two lower panels we show
a few allowed optical transitions indicated by arrows with chemical
potential shown by the horizontal black dashed line. The left panel
is for B=1 Tesla and the right for B=40 Tesla. Here
$\alpha/\hbar=4.3\times10^{5}$ m/s representative of $Bi_{2}Te_{3}$.
For $Bi_{2}Se_{3}$ it is instead $5\times10^{5}$ m/s. } \label{fig3}
\end{figure}

\begin{figure}[tp]
\begin{center}
\includegraphics[height=3in,width=3.0in]{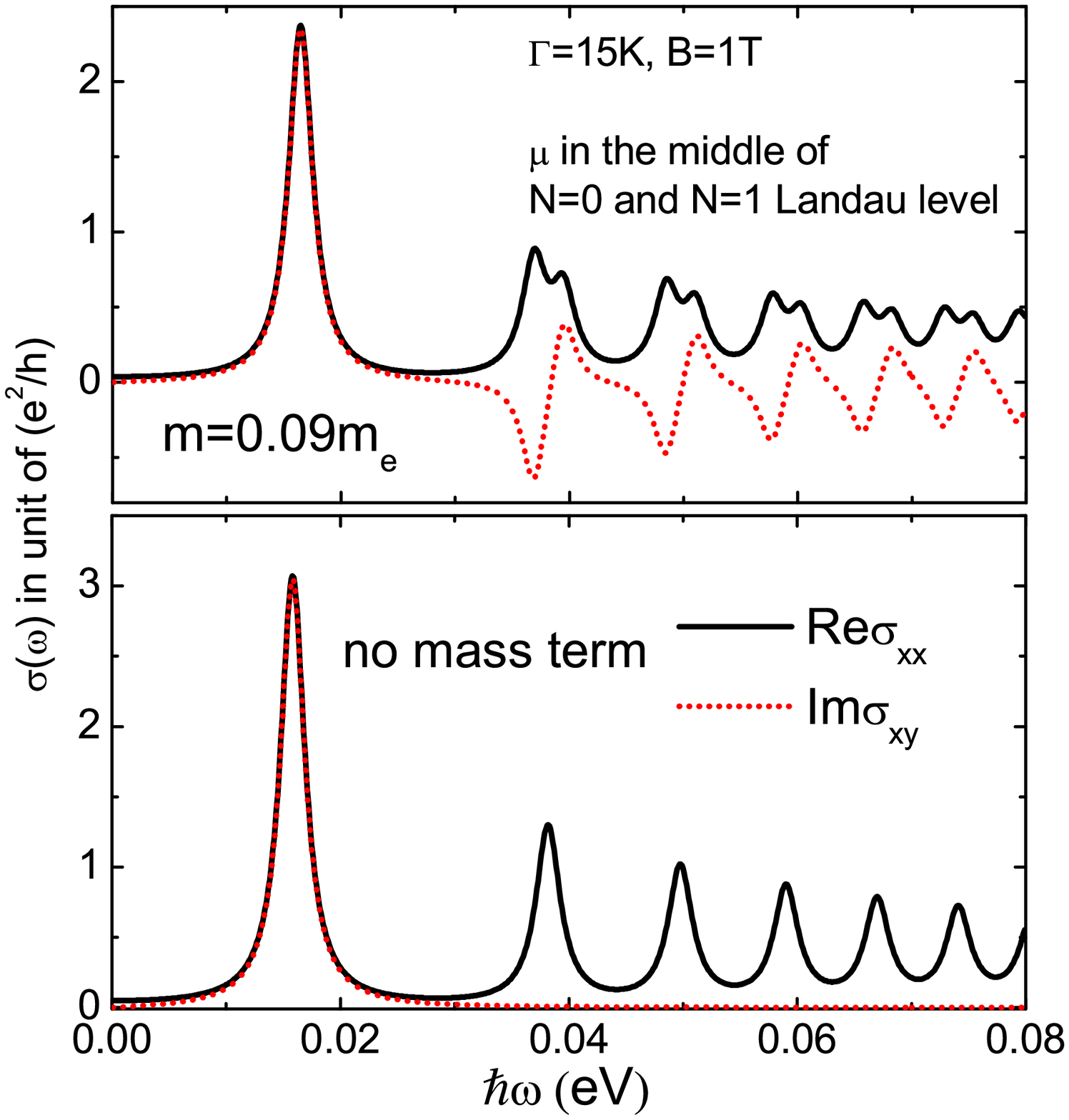}
\includegraphics[height=3in,width=3.0in]{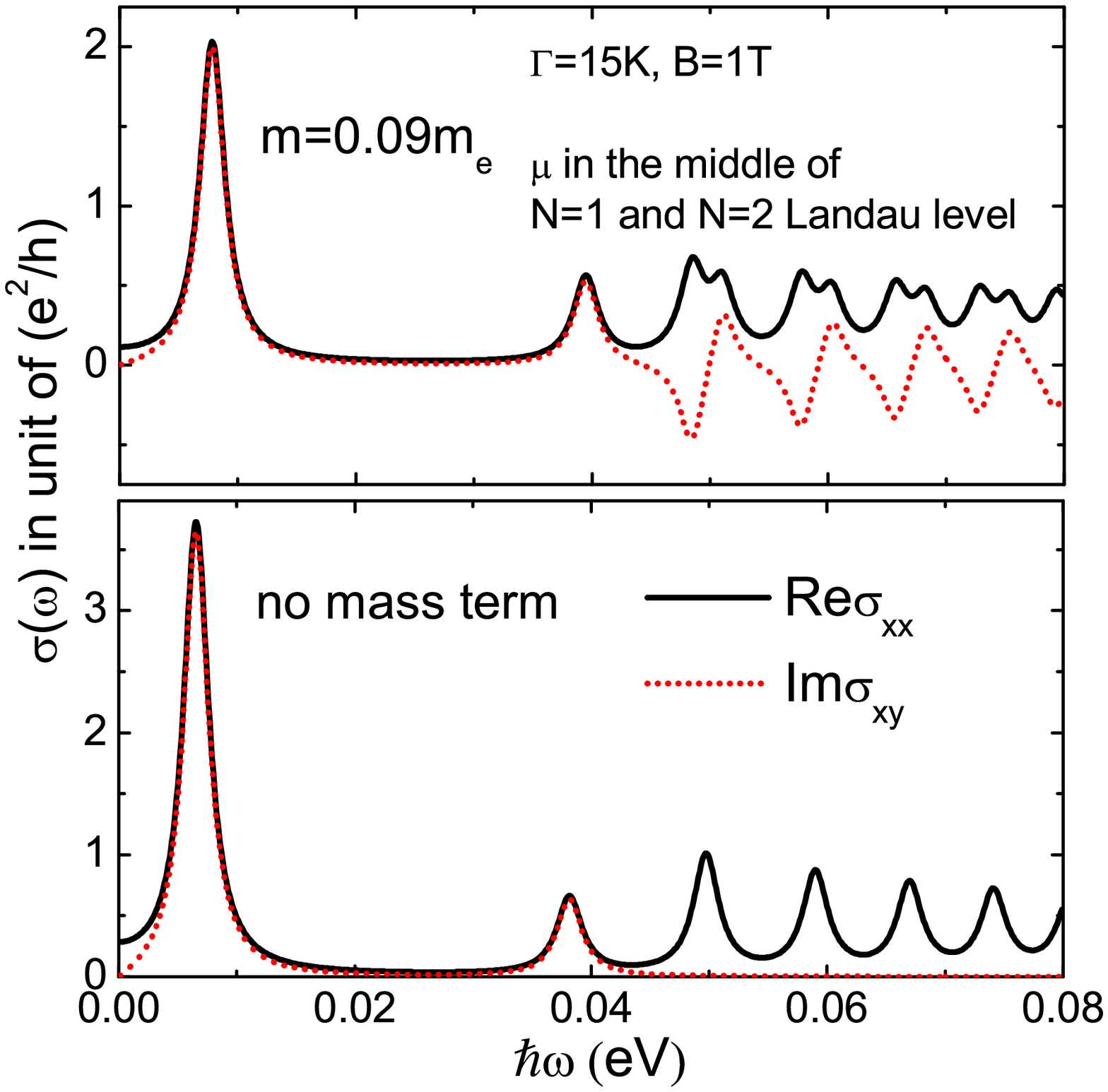}
\end{center}
\caption{(Color online) The real part of the longitudinal optical
conductivity $\sigma_{xx}(\omega)$ (solid black curve) compared with
the imaginary part of the transverse Hall conductivity
$\sigma_{xy}(\omega)$ (dotted red curve) in units of $e^{2}/h$ as a
function of photon energy $\hbar\omega$ in eV. The electron mass
term has $m=0.09m_{e}$, the residual scattering rate is $\Gamma=15K$
and the magnetic field B=1 Tesla. The top two frames are for
chemical potential falling between N=0 and N=1 Landau level while in
the lower two frames it falls between N=1 and N=2. In each case we
show first results for finite m and below for $m=\infty$ (pure
Dirac) which is included for comparison. Here
$\alpha/\hbar=4.3\times10^{5}$ m/s.} \label{fig4}
\end{figure}

\begin{figure}[tp]
\begin{center}
\includegraphics[height=3.0in,width=3.0in]{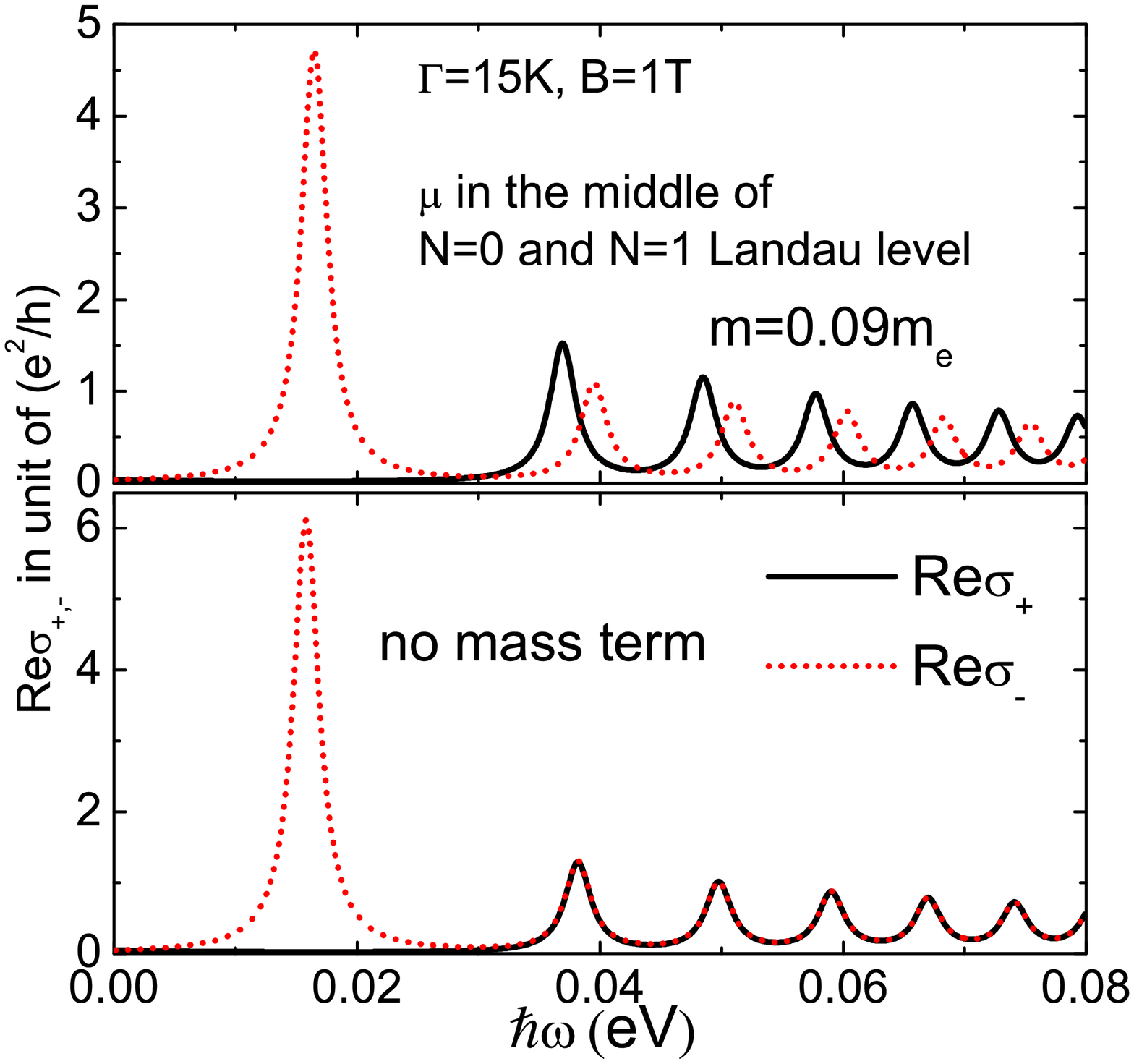}
\includegraphics[height=3.0in,width=3.0in]{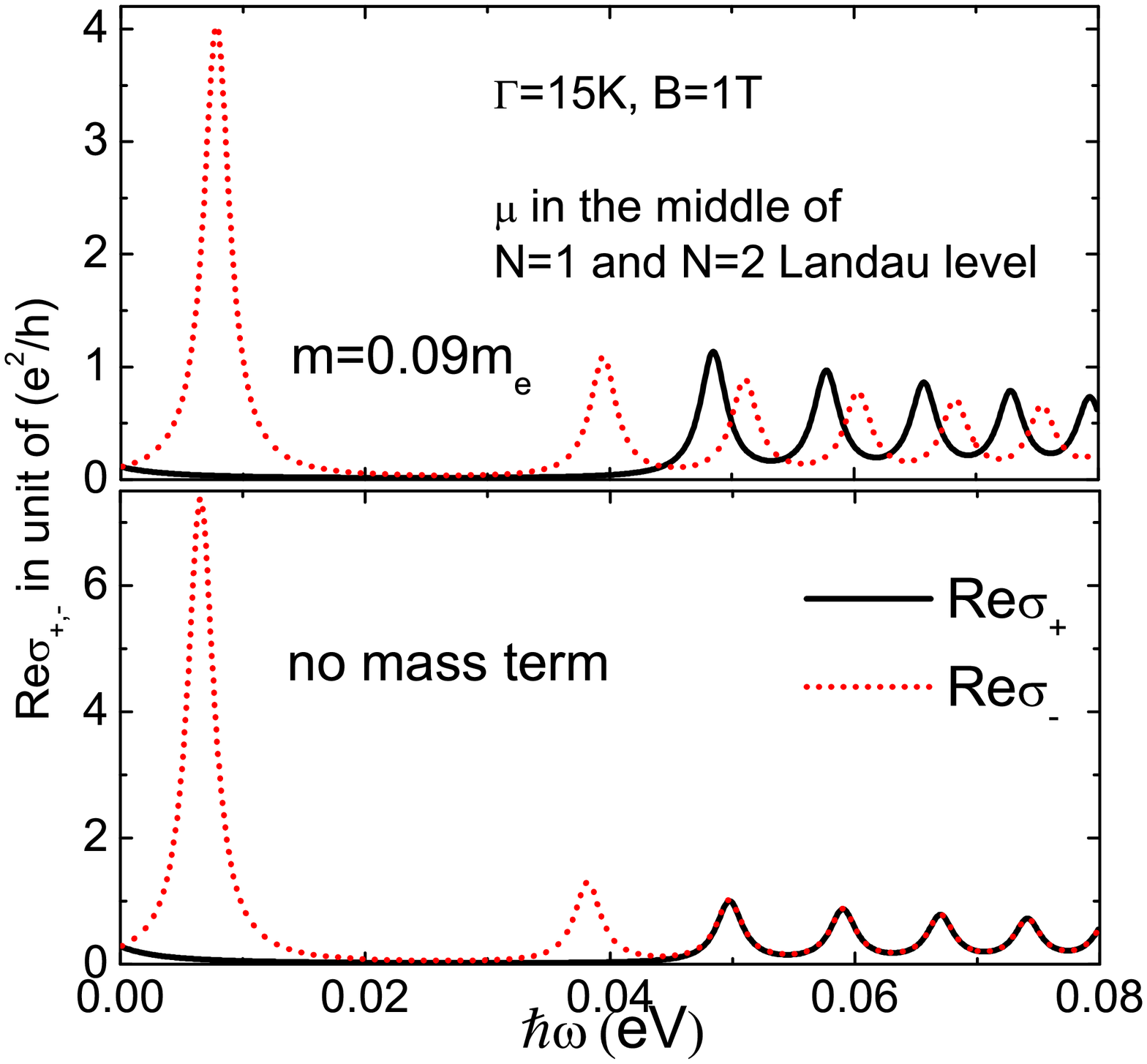}
\end{center}
\caption{(Color online) The real part of right (+) and left handed
circularly polarized optical conductivity $\sigma_{\pm}(%
\omega)$ in units of $e^{2}/h$ as a function of photon energy $\hbar%
\omega$ in eV. In all panels the residual scattering rate
$\Gamma=15K$ and the magnitude of the magnetic field is set at B=1
Tesla. The top two panels are for a case when the chemical potential
falls between N=0 and N=1 Landau level and for the bottom two it
falls between N=1 and N=2. In all cases solid black is for right
hand polarization and the dotted red for left polarization. The
panels come in pairs; in the top frame the quadratic term in the
Hamiltonian has mass $m=0.09m_{e}$ while $m=\infty$ (pure Dirac) for
the bottom frame of each pair and is for comparison. Here
$\alpha/\hbar=4.3\times10^{5}$ m/s.} \label{fig5}
\end{figure}

\begin{figure}[tp]
\begin{center}
\includegraphics[height=3.0in,width=3.0in]{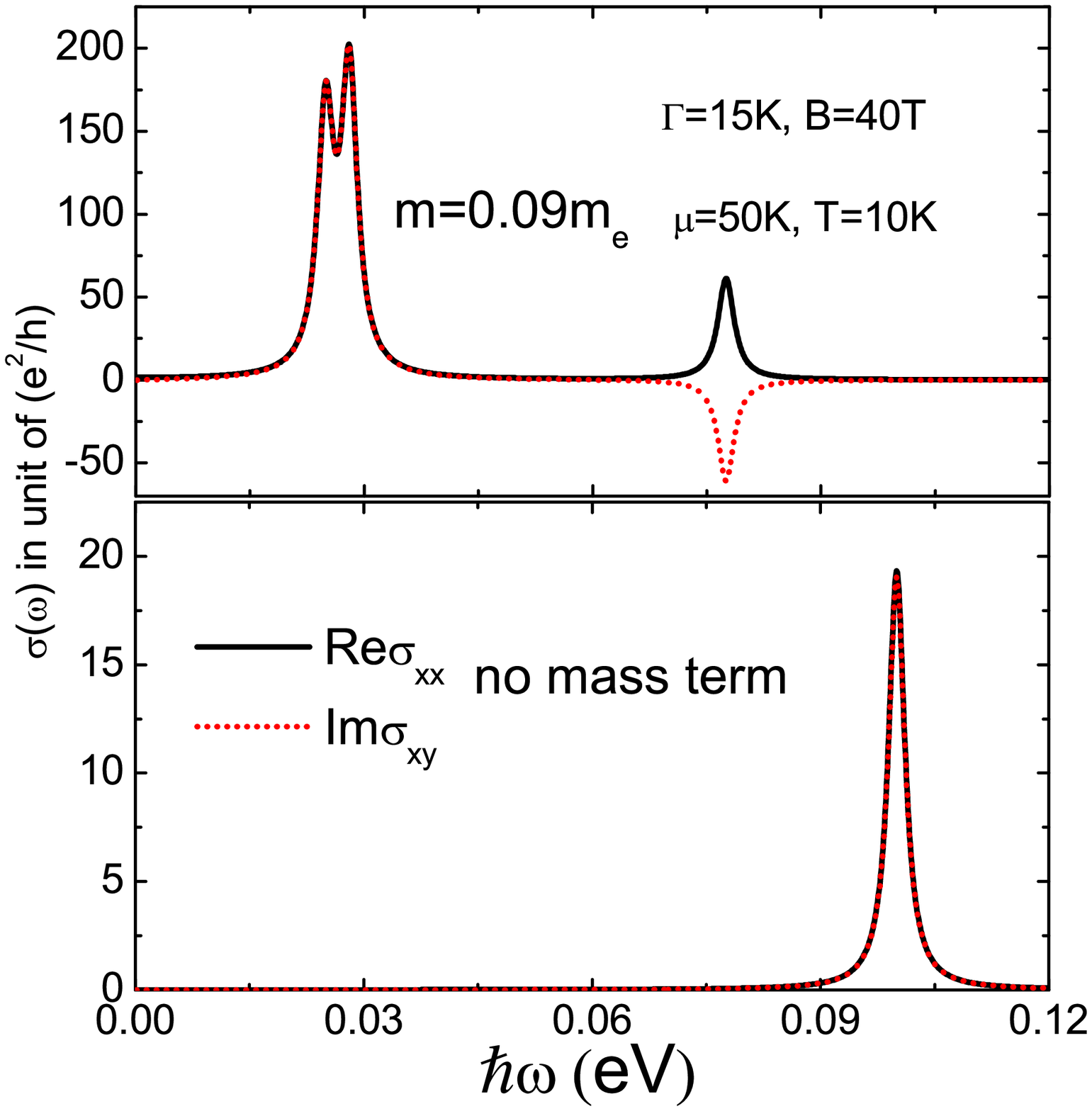}
\includegraphics[height=3.0in,width=3.0in]{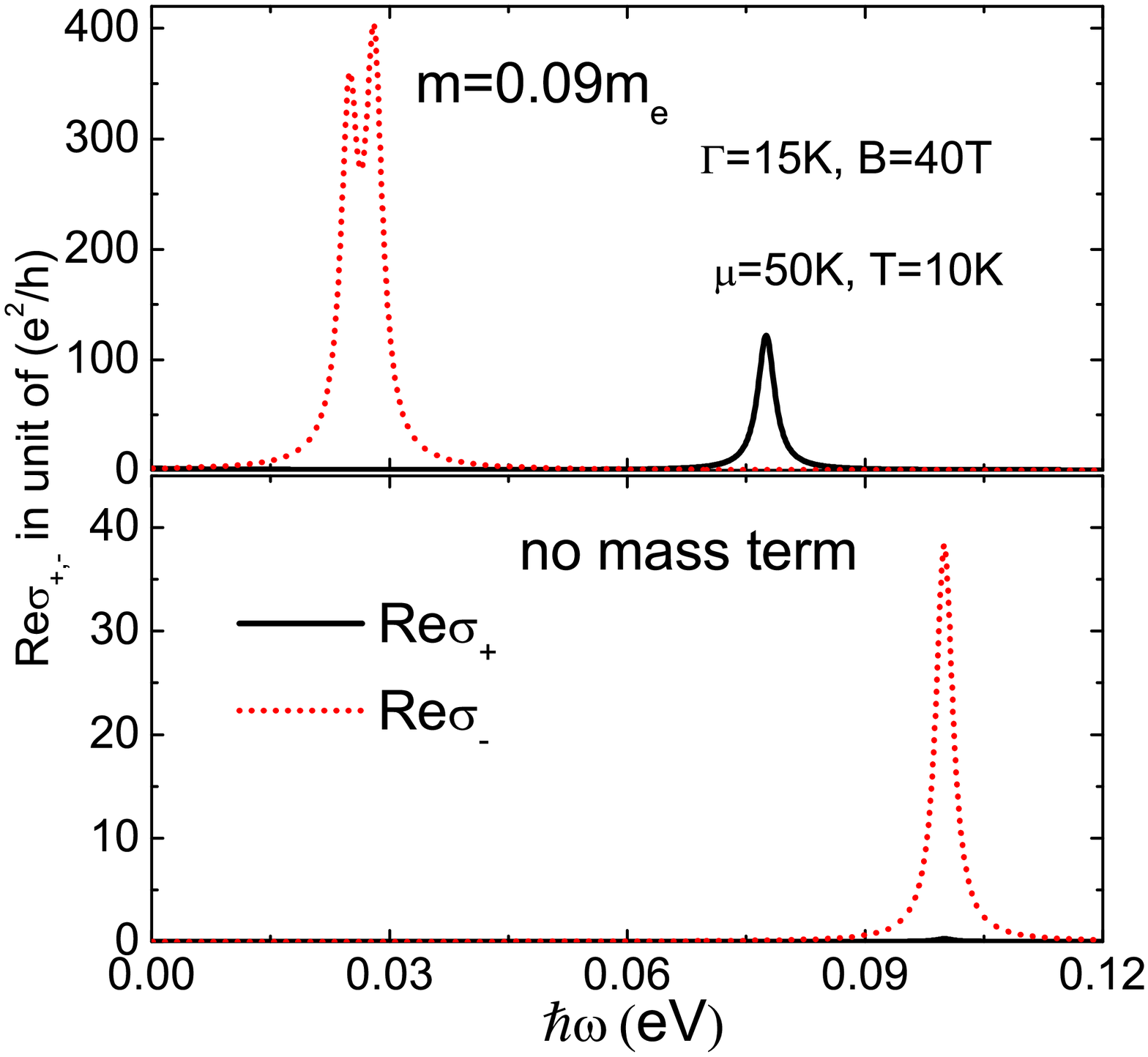}
\end{center}
\caption{(Color online) The top pair of frames give the real part
(imaginary
part) of the longitudinal (transverse Hall) conductivity $Re\sigma%
_{xx}(\omega)$, solid black ($Im\sigma_{xy}(\omega)$%
, red dotted) in unit of $e^{2}/h$ as a function of photon energy $\hbar%
\omega$ in eV. The top panel has a finite quadratic piece in its
dispersion curve with mass $m=0.09m_{e}$ while the bottom panel has
$m=\infty$. In
both cases the residual scattering rate $\Gamma=15K$ the chemical potential $%
\mu=50K$, the temperature T=10K and the magnetic field is set at
B=40 Tesla. The bottom two frames are for the same parameter set and
the
notation is the same. What is plotted however is the right $\sigma%
_{+}$ solid black and left $\sigma_{-}$ dotted red circularly
polarized conductivity. Here $\alpha/\hbar=4.3\times10^{5}$ m/s. }
\label{fig6}
\end{figure}

\begin{figure}[tp]
\begin{center}
\includegraphics[height=3.0in,width=3.0in]{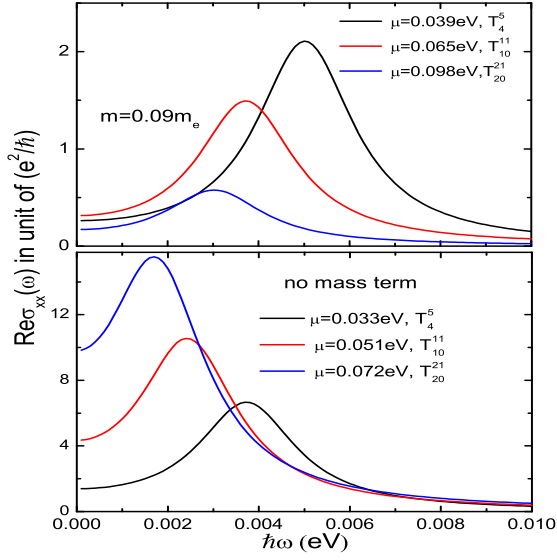}
\end{center}
\caption{(Color online) The semiclassical limit of the real part of
the longitudinal conductivity $Re\sigma_{xx}(\omega)$ in units of
$e^{2}/h$ as a function of photon energy $\hbar\omega$ in eV. The
top and bottom frames give results when $m=0.09m_{e}$ and $m=\infty$
(no quadratic mass term) respectively. Also identified are the
optical transition between Landau levels that are involved (see
lower right panel of Fig.~\ref{fig3}) and the corresponding values
of chemical potential $\mu$. The three cases considered are color
coded. Note that for $m=\infty$ (lower panel) the amplitude of the
conductivity at the cyclotron frequency increases monotonically as
the chemical potential increases while instead it decreases when
$m=0.09m_{e}$. Here $\alpha/\hbar=4.3\times10^{5}$ m/s.}
\label{fig7}
\end{figure}

\begin{figure}[tp]
\begin{center}
\includegraphics[height=3.0in,width=3.0in]{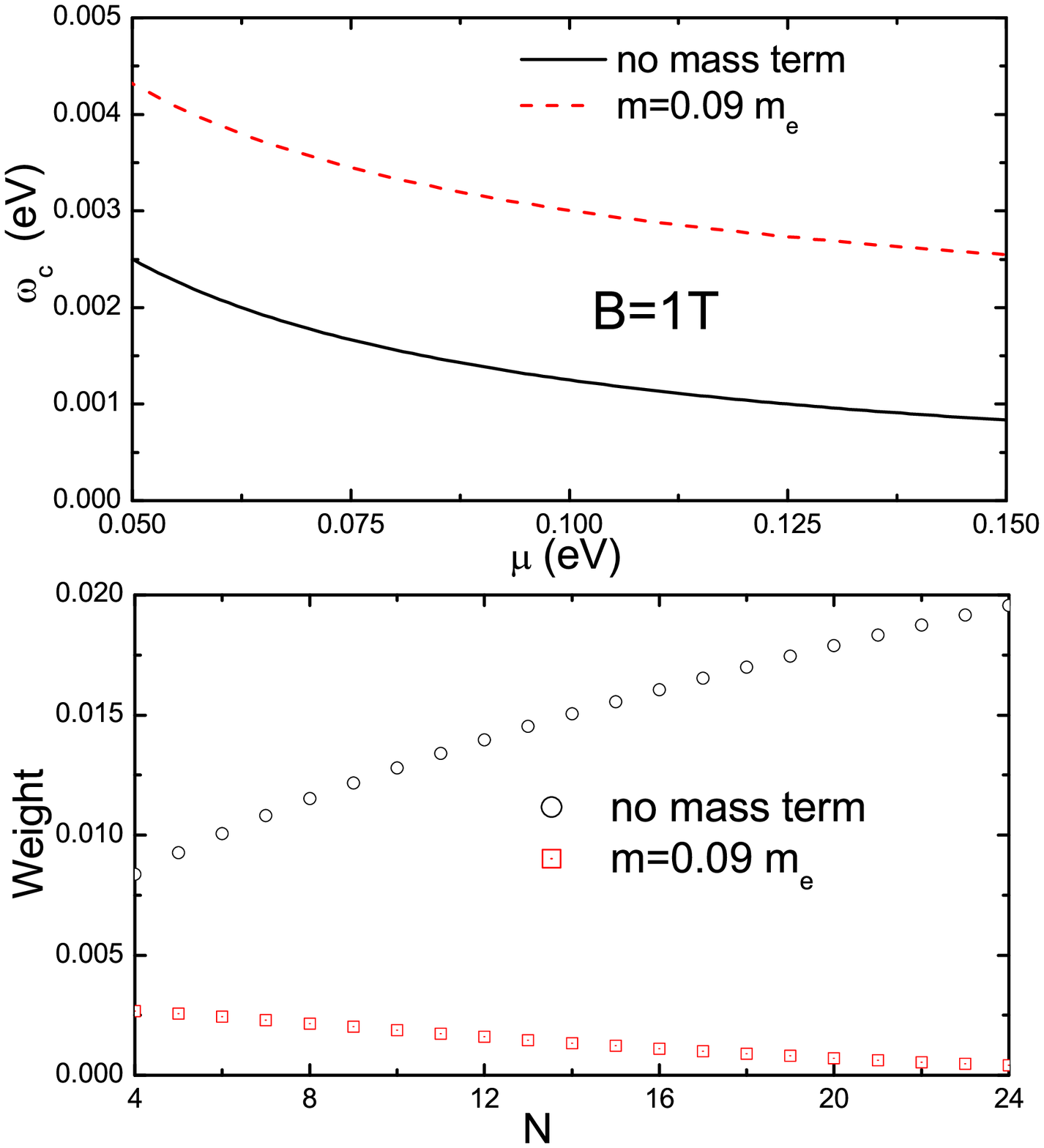}
\end{center}
\caption{(Color online)Top panel gives the semiclassical cyclotron
frequency as a function of chemical potential $\mu$ and the bottom
panel gives the optical spectral weight under the $N$'th line for
$m=0.09m_{e}$ (open square) compared with its value for pure Dirac
(open circle). Here $\alpha/\hbar=4.3\times10^{5}$ m/s.}
\label{fig8}
\end{figure}

We begin with the Hamiltonian for fermions on the surface of a
topological insulator of the form
\begin{equation}
H_{0}=\frac{\hbar ^{2}k^{2}}{2m}+\hbar v_{F}(k_{x}\sigma _{y}-k_{y}\sigma
_{x})  \label{H0}
\end{equation}%
where the first term gives the kinetic energy of a fermion of mass
$m$ with quadratic (Schr\"{o}dinger) in momentum (k) dispersion
curve and the second piece is a term which describes relativistic
Dirac fermions with velocity $v_{F}$. In Eq.~(\ref{H0}) $\sigma
_{x}$ and $\sigma _{y}$ are Pauli spin matrices. In a magnetic field
$B$ oriented perpendicular to the plane of the fermion motion the
Hamiltonian becomes
\begin{eqnarray}
H_{0} &=&\frac{\hbar ^{2}|-i\mathbf{\nabla }+e\mathbf{A}/\hbar |^{2}}{2m}
\notag \\
&&+\alpha \lbrack (-i\partial _{x}+eA_{x}/\hbar )\sigma _{y}-(-i\partial
_{y}+eA_{y}/\hbar )\sigma _{x}]
\end{eqnarray}%
where $\alpha =\hbar v_{F}$ and $\mathbf{A}$ is the vector potential. In the
Landau gauge $\mathbf{A\equiv (0,}B\hat{x}\mathbf{)}$ and we get%
\begin{eqnarray}
H_{0} &=&\frac{\hbar ^{2}[(-i\partial _{x})^{2}+(-i\partial _{y}+eB\hat{x}%
/\hbar )^{2}]}{2m}  \notag \\
&&+\alpha \lbrack (-i\partial _{x})\sigma _{y}-(-i\partial _{y}+eB\hat{x}%
/\hbar )\sigma _{x}]  \label{HM}
\end{eqnarray}%
Raising and lowering operators
\begin{eqnarray}
a^{\dag } &=&l_{B}/\sqrt{2}[-\partial _{x}+(x+x_{0})/l_{B}^{2}],  \notag \\
a &=&l_{B}/\sqrt{2}[\partial _{x}+(x+x_{0})/l_{B}^{2}]
\end{eqnarray}%
with the magnetic length $l_{B}=1/\sqrt{e|B|/\hbar }$ and $%
x_{0}=k_{y}l_{B}^{2}$ can be used to reduce Eq.~(\ref{HM}) to the form%
\begin{equation}
H_{0}=\frac{\hbar ^{2}[a^{\dag }a+1/2]}{ml_{B}^{2}}-\sqrt{2}\alpha
/l_{B}[
\begin{array}{cc}
0 & a \\
a^{\dag } & 0
\end{array}
]
\end{equation}
If a Zeeman splitting term ($\Delta $) is added, the Hamiltonian of
interest becomes
\begin{equation}
H_{0}=[%
\begin{array}{cc}
\hbar ^{2}[a^{\dag }a+1/2]/ml_{B}^{2}+\Delta & (\sqrt{2}\alpha /l_{B})a \\
(\sqrt{2}\alpha /l_{B})a^{\dag } & \hbar ^{2}[a^{\dag
}a+1/2]/ml_{B}^{2}-\Delta \label{HF}%
\end{array}%
]
\end{equation}%
The eigenstates of Eq.~(\ref{HF}) are a mixture of the spin up ($N-1$)'th
Landau level ($|N-1\rangle _{\uparrow }$) and spin down $N$'th Landau level (%
$|N\rangle _{\downarrow }$) which we denote by%
\begin{equation}
|N,s\rangle =\left[
\begin{array}{c}
C_{\uparrow ,N,s}|N-1\rangle _{\uparrow }\notag \\
C_{\downarrow ,N,s}|N\rangle _{\downarrow }%
\end{array}%
\right]
\end{equation}%
with $s=+/-$ corresponding to positive/negative energy states. The eigen
energies are%
\begin{equation}
E_{N,s}=\hbar ^{2}N/(ml_{B}^{2})+s\sqrt{[\hbar
^{2}/(2ml_{B}^{2})]^{2}+2N\alpha ^{2}/l_{B}^{2}}  \label{landau}
\end{equation}%
For $N>0$ and for $N=0$
\begin{equation}
E_{N=0}=\hbar ^{2}/(2ml_{B}^{2})
\end{equation}%
where for simplicity we have set the Zeeman splitting to zero. The
corresponding eigenfunctions are obtained from the solutions of the equation%
\begin{equation}
\left( -E_{0}/2-s\sqrt{E_{0}^{2}/4+2NE_{1}^{2}}\right) C_{\uparrow ,N,s}+%
\sqrt{2NE_{1}^{2}}C_{\downarrow ,N,s}=0\label{sl}
\end{equation}%
where we have introduced the shorthand notation $E_{0}=\hbar e|B|/(m)$ and $%
E_{1}=\frac{\alpha }{l_{B}}$ which refer respectively to the kinetic
energy part that originates from the quadratic part
(Schr\"{o}dinger) and the linear (Dirac) part of the original
Hamiltonian [Eq.~(\ref{H0})]. We can introduce a measure of
"Diracness" $P$ as the ratio $E_{1}^{2}/E_{0}^{2}$. When
$P\rightarrow \infty $ the system is pure Dirac and as $P$ decreases
the system acquires more and more of a Schr\"{o}dinger character.
For $N>0$ the
solution of Eq.~(\ref{sl}) is%
\begin{eqnarray}
C_{\uparrow ,N,s} &=&\frac{\sqrt{\sqrt{1/4+2NP}-s/2}}{J}  \notag \\
C_{\downarrow ,N,s} &=&\frac{s\sqrt{\sqrt{1/4+2NP}+s/2}}{J}
\end{eqnarray}%
with $J=\sqrt{2\sqrt{1/4+2NP}}$. For the special case $N=0$,
$C_{\uparrow ,0}=0$ and $C_{\downarrow ,0}=1$ and only $s=+$ need be
considered. It is interesting to make an estimate of $P$ for
specific topological insulators. Following the work in
Ref.[\onlinecite{Liu}] we obtain for $Bi_{2}Te_{3}$ $m\alpha
^{2}/2\hbar ^{2}=0.048eV$ and $eBv_{F}^{2}\hbar =1.2\times 10^{-4}$ $%
(eV)^{2} $ for a magnetic field of one Tesla and for $Bi_{2}Se_{3}$ we get $%
0.115eV$ and $1.6\times 10^{-4}$ $(eV)^{2}$ respectively. In terms of the
bare electron mass $m_{e}$ we have $m/m_{e}=0.09$ and $0.16$ for $%
Bi_{2}Te_{3}$ and $Bi_{2}Se_{3}$ respectively. For $Bi_{2}Te_{3}$
the "Diracness" parameter $P=74.2$ for an external magnetic field of
$1$ Tesla. The energy $E_{N,s}\cong E_{1}\{N/\sqrt{P}+s\sqrt{2N}\}$
($N\neq 0$) with the first term representative of the leading
contribution to the energy coming from the quadratic piece of the
Hamiltonian and the second term comes from the Dirac piece. The
Schr\"{o}dinger piece provides an $8\%$ correction for $N=1$ ($s=1
$) while this rises to $26\%$ for $N=10$. Because $P$ goes like the
inverse of $B$, for $B=40$ Tesla $P=1.86$ and the correction to the
$N=1$ case becomes $52\%$.

The matrix Green's function associated with our Hamiltonian is given by%
\begin{equation}
\hat{G}_{0}(N,i\omega _{n})=\frac{1}{2}\sum_{s=\pm }(1+s\mathbf{F}_{\mathbf{k%
}}\cdot \mathbf{\sigma })G_{0}(N,s,i\omega _{n})
\end{equation}%
with%
\begin{equation}
\mathbf{F}_{\mathbf{k}}=\frac{(\sqrt{2N}E_{1},0,-E_{0}/2)}{\sqrt{%
E_{0}^{2}/4+(\sqrt{2N}E_{1})^{2}}}
\end{equation}%
and%
\begin{eqnarray}
&&G_{0}(N,s,i\omega _{n}\rightarrow \omega +i\delta )  \notag \\
&=&\frac{1}{\omega +\mu -NE_{0}-s\sqrt{E_{0}^{2}/4+(\sqrt{2N}E_{1})^{2}}%
+i\delta } \label{G0}
\end{eqnarray}
from which we can compute the density of state $D(\omega )$ given by
\begin{equation}
D(\omega )=\frac{-1}{\pi }\frac{1}{2\pi l_{B}^{2}}[\sum_{N=1,s=\pm
}ImG_{0}(N,s,\omega )+ImG_{0}(0,+,\omega )]
\end{equation}
Without a magnetic field we would get%
\begin{equation}
D(\omega ,B=0)=\frac{-1}{\pi }\frac{1}{2\pi }\int_{0}^{k_{cut}}kdk\sum_{s=%
\pm }ImG_{0}(k,s,\omega )
\end{equation}
with%
\begin{equation}
G_{0}(k,s,\omega )=\frac{1}{\omega +\mu -\hbar
^{2}k^{2}/(2m)-s\alpha k+i\delta }\label{G00}
\end{equation}%
and $\mu$ in Eq.~(\ref{G0}) and Eq.~(\ref{G00}) is the chemical
potential where the momentum cut off is taken as
$k_{cut}=\frac{\alpha m}{\hbar ^{2}}$. Fig.~\ref{fig1} is a
schematic of the fermion dispersion relation in our model displaying
the 2-D Brillouin zone (BZ) of the surface states and the hourglass
structure with Dirac point at the $\Gamma $ point in the BZ.
Fig.~\ref{fig2} gives results for the density of states $D(\omega )$
as a function of energy $\hbar \omega $ in eV. The top frame
contains a quadratic contribution with $m=0.09m_{e}$ (our estimate
for $Bi_{2}Te_{3}$) while the bottom frame which is included for
comparison has $m=\infty $, i.e. represents the pure relativistic
Dirac limit. In both cases the magnetic field was set to $1$ Tesla
in the red dotted curve which is to be compared with the continuous
black curve for $B=0$. Comparing first the black curves in top and
bottom frame we see that including the quadratic part in the
Hamiltonian [Eq.~(\ref{H0})] has a drastic effect on the density of
state.
For pure Dirac there is particle-hole symmetry about the Dirac point at $%
\omega =0$ while this no longer is the case in the top frame. Both
positive and negative energy regions are changed with the largest
difference seen in the negative energy region. This results directly
from the hourglass nature of the dispersion curves seen in the
schematic of Fig.~\ref{fig1}. This change in geometry of the
dispersion curves also has a direct effect on the oscillations seen
in the red dotted curves. The difference between top and bottom
frame are particularly large at negative energies as the large peak
around $0.05eV$ in the $B=0$ density of states is sampled. But there
are also very significant changes at positive energies both in the
position of the peaks and in their amplitude. While in the lower
frame (pure Dirac) the red dotted curve around $\hbar \omega
=0.08eV$ has nearly merged with the solid black curve for $B=0$,
this is not the case in the upper curve where the amplitude of the
Landau level oscillations remain very significant.

\section{Magneto-optical conductivity}
The magneto-optical conductivity $\sigma _{\alpha \beta }(\omega )$
based on a system of Landau levels in the clean limit is given by
the standard formula
\begin{eqnarray}
\sigma _{\alpha \beta }(\omega ) &=&\frac{-i}{2\pi l_{B}^{2}}%
\sum_{N,N^{\prime },s,s^{\prime }}\frac{f_{N,s}-f_{N^{\prime },s^{\prime }}}{%
E_{N,s}-E_{N^{\prime },s^{\prime }}}  \notag \\
&&\times \frac{\langle N,s|j_{\alpha }|N^{\prime },s^{\prime }\rangle
\langle N^{\prime },s^{\prime }|j_{\beta }|N,s\rangle }{\omega
-E_{N,s}+E_{N^{\prime },s^{\prime }}+i/(2\tau )}  \label{CAB}
\end{eqnarray}%
where the current operator $j_{\alpha }$ is related to velocity $v_{\alpha }$
and
\begin{eqnarray}
v_{x} &=&\frac{\hbar k_{x}}{m}+\frac{\alpha }{\hbar }\sigma _{y}  \notag \\
&=&i\frac{\hbar }{m}(a^{\dag }-a)/(\sqrt{2}l_{B})+\frac{\alpha }{\hbar }%
\sigma _{y}  \notag \\
v_{y} &=&\frac{\hbar (k_{y}+eA_{y}/\hbar )}{m}-\frac{\alpha }{\hbar }\sigma
_{x}  \notag \\
&=&-\frac{\alpha }{\hbar }\sigma _{x}+\frac{\hbar }{m}(a^{\dag }+a)/(\sqrt{2}%
l_{B})
\end{eqnarray}
In Eq.~(\ref{CAB}) $1/(2\tau )$ is a small residual scattering rate.
Working out the appropriate matrix elements gives for the
longitudinal dynamic conductivity
\begin{eqnarray}
\sigma _{xx}(\omega ) &=&\frac{-ie^{2}}{2\pi l_{B}^{2}}\sum_{s}\frac{%
f_{0}-f_{1,s}}{E_{N=0}-E_{1,s}}Q_{x}(0,1,+,s)  \notag \\
&&\times \langle 0|v_{x}|1,s\rangle \langle 1,s|v_{x}|0\rangle   \notag \\
&&+\frac{-ie^{2}}{2\pi l_{B}^{2}}\sum_{N=1,s,s^{\prime }}\frac{%
f_{N,s}-f_{N+1,s^{\prime }}}{E_{N,s}-E_{N+1,s^{\prime }}}Q_{x}(N,N+1,s,s^{%
\prime })  \notag \\
&&\times \langle N,s|v_{x}|N+1,s^{\prime }\rangle \langle N+1,s^{\prime
}|v_{x}|N,s\rangle   \label{sxx}
\end{eqnarray}%
with $f_{N,s}$ the Fermi-Dirac distribution function given by
$1/(e^{\beta (\omega -\mu )}+1)$ with $\beta $ the inverse
temperature $T$ and $\mu $ the chemical potential. The transverse
dynamic Hall conductivity
\begin{eqnarray}
\sigma _{xy}(\omega ) &=&\frac{-e^{2}}{2\pi l_{B}^{2}}\sum_{s}\frac{%
f_{0}-f_{1,s}}{E_{N=0}-E_{1,s}}Q_{y}(0,1,+,s)  \notag \\
&&\times \langle 0|v_{x}|1,s\rangle \langle 1,s|v_{x}|0\rangle   \notag \\
&&+\frac{-e^{2}}{2\pi l_{B}^{2}}\sum_{N=1,s,s^{\prime }}\frac{%
f_{N,s}-f_{N+1,s^{\prime }}}{E_{N,s}-E_{N+1,s^{\prime }}}Q_{y}(N,N+1,s,s^{%
\prime })  \notag \\
&&\times \langle N,s|v_{x}|N+1,s^{\prime }\rangle \langle N+1,s^{\prime
}|v_{x}|N,s\rangle   \label{sxy}
\end{eqnarray}
where
\begin{eqnarray}
&&\binom{Q_{x}(N,N^{\prime },s,s^{\prime })}{Q_{y}(N,N^{\prime },s,s^{\prime
})}  \notag \\
&=&\frac{1}{\omega +E_{N,s}-E_{N^{\prime},s^{\prime }}+i/(2\tau )}  \notag \\
&&\pm \frac{1}{\omega +E_{N^{\prime},s^{\prime }}-E_{N,s}+i/(2\tau
)}
\end{eqnarray}
Here we will be interested in the absorptive part of the conductivity namely
$Re\sigma _{xx}$ and $Im\sigma _{xy}$ which can be written in the form
\begin{eqnarray}
&&\binom{Re\sigma _{xx}(\omega )}{Im\sigma _{xy}(\omega )}  \notag \\
&=&\mp \frac{e^{2}}{2\hbar }\sum_{s}\frac{f_{0}-f_{1,s}}{E_{N=0}-E_{1,s}}%
\frac{F(0,s)}{H(0,s)}E_{0}  \notag \\
&&\times \lbrack \delta (\hbar \omega -E_{0}H(0,s))\pm \delta (\hbar \omega
+E_{0}H(0,s)]  \notag \\
&&\mp \frac{e^{2}}{2\hbar }\sum_{N=1,s,s^{\prime }}(f_{N,s}-f_{N+1,s^{\prime
}})\frac{F(N,s,s^{\prime })}{H(N,s,s^{\prime })}E_{0}  \notag \\
&&\times \lbrack \delta (\hbar \omega -E_{0}H(N,s,s^{\prime }))  \notag \\
&&\pm \delta (\hbar \omega +E_{0}H(N,s,s^{\prime })] \label{RXY}
\end{eqnarray}
where the delta function $\delta (x)$ conserve energy in the optical
absorption process and we have defined%
\begin{equation}
H(N,s,s^{\prime })=-1+s\sqrt{1/4+2NP}-s^{\prime }\sqrt{1/4+2(N+1)P}
\label{HN}
\end{equation}%
and%
\begin{eqnarray}
F(N,s,s^{\prime }) &=&(\frac{\sqrt{N}}{\sqrt{2}}C_{\uparrow ,N+1,s^{\prime
}}^{\ast }C_{\uparrow ,N,s}+\frac{\sqrt{N+1}}{\sqrt{2}}C_{\downarrow
,N+1,s^{\prime }}^{\ast }  \notag \\
&&\times C_{\downarrow ,N,s}-\sqrt{P}C_{\uparrow ,N+1,s^{\prime }}^{\ast
}C_{\downarrow ,N,s})^{2}  \label{FN}
\end{eqnarray}%
for $N\neq 0$ and for $N=0$
\begin{equation}
H(0,s)=-1/2-s\sqrt{1/4+2P}
\end{equation}%
\begin{equation}
F(0,s)=(\frac{1}{\sqrt{2}}C_{\downarrow ,1,s}-\sqrt{P}C_{\uparrow ,1,s})^{2}
\end{equation}%
In Fig.~\ref{fig3} we show results for the Landau level energies obtained
from Eq.~(\ref{landau}) for the specific case of $Bi_{2}Te_{3}$ with $%
m/m_{e}=0.09$ as a function of magnetic field $B$ in Tesla. The
solid lines are for positive $N$'s as well as $N=0$ while the dotted
are for negative $N$'s. The energies of the positive branch levels
are not as strongly modified by the quadratic Schr\"{o}dinger term
in Eq.~(\ref{H0}) than are the energies of the negative branch. For
the pure Dirac case there would be mirror symmetry between positive
and negative levels and they would all scale as the square root of
the magnetic field $B$. Now the positive $N$ levels show small
deviations from $\sqrt{B}$ law while the negative $N$ levels begin
to bend over upward as $B$ increases and can even cross the $E=0$
axis at $40$ Tesla for the larger values of $N$ shown. The solid
black curve for $N=0$ is linear in $B$ and goes to zero at $B=0$.
For pure Dirac this level would remain at zero for all values of
magnetic field. The lack of perfect mirror symmetry between positive
and negative branches of the spectrum has important
implications for the peaks seen in both $Re\sigma _{xx}(\omega )$ and $%
Im\sigma _{xy}(\omega )$ as a function of $\omega $. This is
illustrated in Fig.~\ref{fig4} which has four frames. The upper two
are for a value of chemical potential which falls between $N=0$ and
$N=1$ Landau levels (LL) while in the two lower frames the chemical
potential $\mu $ falls between $N=1$ and $N=2$ LL. In all cases
$\Gamma =1/(2\tau )$ in Eq.~(\ref{sxx}) and Eq.~(\ref{sxy}) was set
to $15K$ and the magnetic field at one Tesla. The frames come in
pairs, the top panel includes a finite mass for the quadratic term
in the Hamiltonian (1) $m/m_{e}=0.09$ while the lower panel of each
pair is for comparison and has $m/m_{e}=\infty $ (the pure Dirac
limit). Also, the solid continuous black curve is for the real part
of the longitudinal conductivity $Re\sigma _{xx}(\omega )$ v.s.
$\omega $ and the dotted red curve for the imaginary part of the
Hall conductivity $Im\sigma _{xy}(\omega )$ v.s. $\omega $. Starting
with the upper two frames, we note first that including the
quadratic term in the dispersion relation has split the peaks in the
solid curves into two except for the first one. This feature can be
traced directly to the energy level scheme of Fig.~\ref{fig3} and
the lack of perfect symmetry between positive and negative branch.
The arrows in the left hand lower frame connect energy levels for
the allowed optical transitions (the chemical potential $\mu =50K$
is shown as the horizontal black dashed line which falls slightly
above the solid black line associated with the $N=0$ level). Note
that the zeroth Landau level is not quite at $E=0$ because $B$ is
finite (1 Tesla). The optical selection rules allow $N$ to change by
only one. In addition one needs to go from occupied to unoccupied
states through the absorption of the photon. The first peak in
$Re\sigma _{xx}(\omega )$ and $Im\sigma _{xy}(\omega )$ come from
the transition indicated by the shortest arrow on the left from
$N=0$ to $N=1$. There is only one such arrow and consequently only
one peak in the conductivity. However for the second peak two arrows
contribute, that from $N=1$ (negative side) to $N=2$ (positive side)
and from $N=2$ (negative side) to $N=1$ (positive side). For the
pure Dirac case these two arrows would have exactly the same length
and there is only one peak in the conductivity as we see in the
lower frame of Fig.~\ref{fig4} for $\mu $ between $N=0$ and $N=1$.
But when there is a quadratic term in Eq.~(\ref{H0}) the symmetry
between positive and negative branch is no longer observed and the
two arrows in question are of slightly different length. This line
is split into two peaks in the conductivity. The amount of splitting
reflects directly the difference in the absolute value of the energy
between positive and negative branch for the same $N$. Here we have
used $m/m_{e}=0.09$ for $Bi_{2}Te_{3}$. For smaller values of $m$
the observed splitting would increase for fixed Dirac spectrum
($\alpha $). The consequences of this mismatch between energies of
positive and negative energy branches is even more striking for the
Hall than it is for the longitudinal conductivity. The red dotted
curve in the upper frame shows a first negative oscillation and then
a positive peak. For the pure Dirac case these two peaks would have
the same energy and hence cancel out perfectly as we see in the
second frame of Fig.~\ref{fig4}. In that case only the first peak
remains in the Hall conductivity, all higher peaks are missing due
to the cancelation just described. Turning next to the second set of
two frames of Fig.~\ref{fig4} we see a similar pattern but with
significant differences that need to be commented upon. In this case
we have increased the value of the chemical potential so that it
falls between the positive $N=1$ and $N=2$ Landau levels. Firstly
note that the first peak has shifted to lower energy while all
others stay at the same energies as in the upper two frames, but the
intensity of the second peak has been reduced by a factor of $2$ for
the pure Dirac case. Secondly, in the mixed Schr\"{o}dinger-Dirac
case the first negative oscillation in Hall conductivity is absent
as is the lower split peak in the longitudinal conductivity. These
features are easily understood with the help of the energy level
diagram left lower frame of Fig.~\ref{fig3}. Moving the chemical
potential level (dashed line) to fall between $N=1$ and $N=2$ we see
that the transition with the shortest arrow is no longer allowed by
the optical selection rule. The second arrow from $N=1$ (negative
branch) to $N=2$ (positive branch) remains a possible optical
transition while the transition from $N=2$ (negative branch) to
$N=1$ (positive branch) is Pauli blocked and is no longer possible.
The final state is already occupied. There is only one arrow that
can contribute and half the line is lost so the second peak is no
longer split. This is one of our important results.

So far we have described in relation to Fig.~\ref{fig4} only the
interband transitions between negative and positive branches. But
the optical selection rule also allow intraband transitions between
$N$ and $N+1$ of the same branch. To understand how it is that the
lowest peak in both $Re\sigma _{xx}(\omega )$ and $Im\sigma
_{xy}(\omega )$ has shifted to lower energies in the two lower
frames of Fig.~\ref{fig4}, we need to include intraband transitions.
Returning to the level diagram of Fig.~\ref{fig3} lower left frame
we can see that for $\mu $ between $N=1$ and $N=2$ levels an optical
transition from $1$(occupied) to $2$ (unoccupied) is now possible
and this gives the first peak (seen in our Fig.~\ref{fig4} two lower
frames) which is a intraband peak and replaces the first interband
peak (in the upper two frames) which is no longer possible. These
results are based on the simplest Hamiltonian (1) which includes
only a Schrodinger quadratic in momentum term and a Dirac linear in
k contribution. Recently Fu \cite{Fu} found that to understand the
ARPES data in $Bi_{2}Te_{3}$ a cubic hexagonal correction needs to
be added to the Hamiltonian. The role of hexagonal warping played in
optics was further discussed in reference [\onlinecite{Li3}]. It was
found to change the constant universal background provided by the
interband optical transitions to a sloped background which increases
with increasing photon energy. The application of a magnetic field
introduces Landau level (LL) oscillations based on this background.
Consequently, as a first approximation, we expect that including
hexagonal warping with a magnetic field would lead to LL
oscillations which would average out at higher energies to a sloped
rather than a constant background.

The peak structure just described for $Re\sigma _{xx}(\omega )$ and $%
Im\sigma _{xy}(\omega )$ has important implication for the behavior of the
conductivity for right and left handedness polarized light defined as $%
\sigma _{\pm }(\omega )\equiv $ $\sigma _{xx}(\omega )\pm i\sigma
_{xy}(\omega )$. This is shown in Fig.~\ref{fig5} which has a direct
correspondence to the data presented in Fig.~\ref{fig4} and which
also has four frames. The top two are for $\mu $ between $N=0$ and
$N=1$ LL and the lower two for $\mu $ between $N=1$ and $N=2$ LL.
Solid black curve is $Re\sigma _{+}(\omega )$ and dotted red is
$Re\sigma _{-}(\omega )$. Note that there are no split peaks in
these quantities but striking differences between the
Schr\"{o}dinger plus Dirac case and pure Dirac (lower frame of each
pair) remain. For pure Dirac, $Re\sigma _{-}(\omega )$ (red dotted
curve) has a single peak corresponding to the lowest energy peak of
Fig.~\ref{fig4}. This peak is missing in $Re\sigma _{+}(\omega )$
which, however, has all the other peaks seen in Fig.~\ref{fig4}. For
the mixed Schr\"{o}dinger plus Dirac case the situation is similar
with one important difference. The higher energy peaks remain in
$Re\sigma_{xx}(\omega )$ (red dotted curve) but they are displaced
in energy with respect to those in the solid black curve. These
differences between pure Dirac and the case with the existence of a
subdominant Schr\"{o}dinger part to the Hamiltonian [Eq.~(\ref{H0})]
could be used to estimate the magnitude of this second contribution.

So far we have considered only the case of $B=1T$ and have found
differences with the pure Dirac case. These differences can be made
much more dramatic by increasing the magnitude of the external
magnetic field as we show in Fig.~\ref{fig6} for magnetic field B=40
Tesla. As we see in the top frame of Fig.~\ref{fig3} for the Landau
level energies, the $N=0$ level has now moved to $E_{0}\simeq
0.025eV$. While for $N=1,2,3$ the valence band energies remain
negative, they have moved to positive values for $N=4$ and above. In
the right hand lower frame of Fig.~\ref{fig3} we show these various
levels as well as the chemical potential (black dashed curve) which
we take to fall just above the $N=4$ level of the negative branch
(dotted cyan lines). All other levels fall outside the energy range
shown in the Figure. Because we are at finite temperature, the
possible optical transitions are shown as arrows from $4$ to $5$,
from $3$ to $4$ and from $1$ to $0$. The first two are close in
energy and give a split peak in the top frame of Fig.~\ref{fig6} for
$Re\sigma _{xx}(\omega )$ with mass term included. The lower peak in
the split pair which is due to the 3 to 4 transition will disappear
at zero temperature as this transition is now Pauli blocked. The
long arrow gives the second single peak at about $0.07eV$ in the
black curve. This transition also contributes to the imaginary part
of the Hall conductivity (dotted red) which shows a negative peak.
It also displays the same lower energy split peak as does the
longitudinal conductivity. Consequently the left circular
polarization conductivity $Re\sigma _{-}(\omega )$ (red dotted)
shows a low energy split peak of twice the amplitude of its value in
$Re\sigma _{xx}(\omega )$ and no other peak while the right circular
polarization conductivity $Re\sigma _{+}(\omega )$ (black solid)
shows a single positive peak at about $0.07eV$. This is to be
contrasted with the pure Dirac case shown in the lower frame of each
pair of diagrams. In that case there is a single line in the energy
range shown. It is present in $Re\sigma_{xx}(\omega )$ (black solid)
and in $Im\sigma_{xy}(\omega )$ (dotted red) with the same amplitude
as well as in $Re\sigma _{-}(\omega )$ (dotted red) with twice the
amplitude. It does not appear in $Re\sigma_{+}(\omega )$.

\section{Semiclassical limit}
The semiclassical limit of the magneto-conductivity is obtained when
the chemical potential $\mu $ is much larger than the magnetic
energy. In this case for $\mu >0$ only intraband transition are
involved as we are interested in the cyclotron resonance energy
range which is much less than $\mu $. This involves large values of
$N$. Let $\mu $ fall between the $N$'th and $N+1$'th Landau level.
So that
\begin{equation}
\mu =NE_{0}+\sqrt{2NE_{1}^{2}+(E_{0}/2)^{2}}\simeq NE_{0}+\sqrt{2N}E_{1}
\label{chem}
\end{equation}
and the energy of the optical transition from $N$ to $N+1$ is given by
\begin{equation}
\hbar \omega _{c}\simeq E_{N+1}-E_{N}
\end{equation}
which is the semiclassical cyclotron frequency. We can solve
Eq.~(\ref{chem}) to get $N$ in terms of $\mu$ and obtain
\begin{equation}
\sqrt{N}=\frac{E_{1}}{\sqrt{2}E_{0}}[\sqrt{1+2\mu E_{0}/E_{1}^{2}}-1]
\end{equation}%
and so
\begin{equation}
\hbar \omega _{c}=\frac{\hbar eBv_{F}^{2}}{\mu }\frac{\sqrt{1+2\mu \hbar
^{2}/m\alpha ^{2}}[\sqrt{1+2\mu \hbar ^{2}/m\alpha ^{2}}+1]}{2}  \label{semi}
\end{equation}
For $m\rightarrow \infty $ we have the well known results for the
pure Dirac case
\begin{equation}
\hbar \omega _{c}=\frac{eBv_{F}^{2}\hbar }{\mu }
\end{equation}
The first correction for $m$ large but not infinite is
\begin{equation}
\hbar \omega _{c}=\frac{eBv_{F}^{2}\hbar }{\mu }[1+\frac{3}{2}\mu
/(mv_{F}^{2})-\frac{1}{4}\mu ^{2}/(mv_{F}^{2})^{2}]
\end{equation}
The pure Schr\"{o}dinger case is obtained as $\alpha \rightarrow 0$
which gives
\begin{equation}
\hbar \omega _{c}=\frac{\hbar eB}{m}
\end{equation}
and the lowest order correction for $\alpha $ small but not zero, is
\begin{equation}
\hbar \omega _{c}=\frac{\hbar eB}{m}[1+\sqrt{m\alpha ^{2}/2\mu \hbar ^{2}}%
+m\alpha ^{2}/2\mu \hbar ^{2}]
\end{equation}
For $Bi_{2}Te_{3}$ we estimate $m\alpha ^{2}/(2\hbar ^{2})=0.048eV$
and $eBv_{F}^{2}\hbar =1.2\times 10^{-4}(eV)^{2}$ and for
$Bi_{2}Se_{3}$ we have $0.115eV$ and $1.6\times 10^{-4}$ $(eV)^{2}$
respectively. For the first case we show in the upper frame of
Fig.~\ref{fig8} the cyclotron energy as a function of chemical
potential (red dashed curve) which we compare with pure Dirac (black
solid curve). Both show similar variation with $\mu $ (inverse $\mu
$ law) but the two curves are significantly displaced from each
other for the case $ m/m_{e}=0.09$. For smaller values of the mass
$m$ the effect of a subdominant Schr\"{o}dinger contribution to the
cyclotron frequency will be even larger.

The real part of the longitudinal conductivity $Re\sigma
_{xx}(\omega )$ v.s. $\omega $ in units of $e^{2}/\hbar $ in the
semiclassical regime is shown in Fig.~\ref{fig7}. The top frame is
for the parameters that we have associated with $Bi_{2}Te_{3}$ with
$m/m_{e}=0.09$ while the bottom frame results when the
Schr\"{o}dinger piece of the energy [Eq.~(\ref{H0})] is dropped and
is for comparison. The values of chemical potential used and the
optical transitions involved are labeled and color coded in the
figure. Not only is the value of the central frequency ($\omega
_{c}$) giving the peak in these line-shapes different when $m/m_{e}$
is finite as compared with its value when $m/m_{e}$ is infinite, but
also the line-shapes themselves are quite different. As $\omega
_{c}$ increases the optical spectral weight (area under the curve
for $Re\sigma _{xx}(\omega )$) decreases in the $m=\infty $ case
while it increases for $m/m_{e}=0.09$. This is illustrated further
in the lower frame of Fig.~\ref{fig8} where we plot the spectral
weight as a function of $N$ with (open squares) and without (open
circles) a mass term. Note that this quantity is related to the
ratio $F(N,+,+)E_{0}/H(N,+,+)$ which appears in Eq.~(\ref{RXY}) with
$H$ and $F$ given by Eq.~(\ref{HN}) and Eq.~(\ref{FN}) respectively.
We see that including a small mass term has a profound effect on
this spectral weight which is another important result of this
analysis.

\section{Summary and Conclusions}
We have studied how adding a small subdominant quadratic in momentum
term to a dominant Dirac dispersion modifies the
magneto-conductivity when Landau levels are formed in a topological
insulator by application of a magnetic field $B$. In such a case the
energies of the Landau levels in conduction and valence band no
longer mirror each other. This means that interband optical
transitions from level $N$ in the valence band to $N+1$ in the
conduction band no longer have the same energy as those from $N+1$
to $N$ and this splits the corresponding absorption line for the
real part of the longitudinal conductivity into two, each carrying
the identical optical spectral weight. The energy of the splitting
is related to the mismatch in energy levels between conduction and
valence band. A similar splitting is found for the imaginary part of
the Hall conductivity. However, for the absorption of circularly
polarized light single peak structures are recovered but in this
case there is a shift in the energy position of the lines between
right and left polarization in contrast to what is found when the
mass term in the electron dispersion curves is zero for pure Dirac.
The semiclassical limit is also affected by the presence of a
subdominant quadratic term. This significantly shifts the value of
the cyclotron frequency away from its pure Dirac value of
$eBv^{2}_{F}\hbar/\mu$ and introduces a more complicated dependence
on chemical potential given in Eq.~(\ref{semi}). The line-shape
associated with the cyclotron resonance is significantly changed.
The optical spectral weight under these curves is found to decrease
with increasing value of the chemical potential rather than increase
as would be the case in a pure relativistic Dirac system.

\begin{acknowledgments}
This work was supported by the Natural Sciences and Engineering
Research Council of Canada (NSERC) and the Canadian Institute for
Advanced Research (CIFAR) and in part by Perimeter Institute for
Theoretical Physics, which is supported by the Government of Canada
through Industry Canada and by the Province of Ontario through the
Ministry of Economic Development and Innovation.
\end{acknowledgments}


\section*{References}

\begin{thebibliography}{99}
\bibitem{Hasan} M. Z. Hasan and C. L. Kane, Rev. Mod. Phys. \textbf{82},
3045 (2010).

\bibitem{Qi1} X.-L. Qi and S.-C. Zhang, Rev. Mod. Phys. \textbf{83}, 1057
(2011).

\bibitem{Hsieh1} D. Hsieh et.al, Nature (London),\textbf{452}, 970 (2008).

\bibitem{Chen} Y. L. Chen, J. G Analytis et.al, Science \textbf{325},
178(2009).

\bibitem{Hsieh2} D. Hsieh et.al, Nature (London),\textbf{460}, 1101 (2009).

\bibitem{Jozwiak} C. Jozwiak, X. L. Chen et.al, Phys. Rev. B \textbf{84},
165113 (2011).

\bibitem{Xu} S.-Y. Xu, X Xia et.al, Science \textbf{332}, 560 (2011).

\bibitem{Qi2} X.-L. Qi and S.-C. Zhang, Physics Today \textbf{63}, 33 (2010).

\bibitem{Hancock} J. N. Hancock, et.al, Phys. Rev. Lett. \textbf{107},
136803(2011).

\bibitem{Li3} Zhou Li and J. P. Carbotte, Phys. Rev. B \textbf{87},
155416 (2013).

\bibitem{Carbotte1} V. P. Gusynin, S. G. Sharapov and J. P. Carbotte, Phys.
Rev. Lett. \textbf{98}, 157402 (2007).

\bibitem{Carbotte2} V. P. Gusynin, S. G. Sharapov and J. P. Carbotte, New J.
Phys. \textbf{11}, 095013 (2009).

\bibitem{Li} Z. Li, E. A. Henniksen et.al, Nature Phys. \textbf{4}, 532
(2008).

\bibitem{Onlita} M. Orlita and M. Potemski, Semicond. Sci. Technol. \textbf{%
25}, 063001 (2010).

\bibitem{Crasee} I. Crasee, J. Levallois et.al, Nature. Phys. \textbf{7}, 48
(2011).

\bibitem{Nicol} E. J. Nicol and J. P. Carbotte, Phys. Rev. B \textbf{77},
155409 (2008).

\bibitem{Stauber} T. Stauber and N. M. R. Peres, J. Phys. Condens. Matter,
\textbf{20}, 055002 (2008).

\bibitem{Li12} Zhou Li and J. P. Carbotte, Phys. Rev. B \textbf{86}, 205425 (2012).

\bibitem{Li13} Zhou Li and J. P. Carbotte, Physica B (2013), http://dx.doi.org/10.1016/j.physb.2013.04.030

\bibitem{Stille} L. Stille, C. J. Tabert and E. J. Nicol, Phys. Rev. B.
\textbf{86}, 195405 (2012).

\bibitem{Carbotte3} V. P. Gusynin, S. G. Sharapov and J. P. Carbotte, J.
Phys. Condens. Matter \textbf{19}, 026222 (2007).

\bibitem{Pound} A. Pound, J. P. Carbotte and E. J. Nicol, Phys. Rev. B
\textbf{85}, 125422 (2012).

\bibitem{Sadowski} M. L. Sadowski, G. Martinez, M. Potemski, C. Berger and W. A. de Heer, Phys.
Rev. Lett. \textbf{97}, 266405 (2006).

\bibitem{Jiang} Z. Jiang et.al, Phys. Rev. Lett. \textbf{98}, 197403 (2007).

\bibitem{Tabert} C. J. Tabert and E. J. Nicol, Phys. Rev. Lett.
\textbf{110}, 197402 (2013).

\bibitem{Schafga} A. A. Schafgans et.al, Phys. Rev. B. \textbf{85}, 195440
(2012).

\bibitem{Ando}  A. A. Taskin and Y. Ando, Phys. Rev. B
\textbf{84}, 035301 (2011).

\bibitem{Liu} C.-X. Liu, X.-L. Qi, H. J. Zhang, X. Dai, Z. Fang and S.-C.
Zhang, Phys. Rev. B \textbf{82}, 045122 (2010).

\bibitem{Fu} L. Fu, Phys. Rev. Lett \textbf{103}, 266801 (2009).

\end{thebibliography}
\end{document}